\documentclass[aps,preprint,showpacs,prd,nofootinbib]{revtex4-1}
\usepackage{amssymb,graphics,graphicx,amsmath}
\newcommand{\be}{\begin{equation}} 
\newcommand{\ee}{\end{equation}}

\begin{document}
\title
{\Large \bf Dark Matter from Binary Tetrahedral Flavor Symmetry}

\author{David A. Eby\footnote{daeby@physics.unc.edu} 
and  Paul H. Frampton\footnote{frampton@physics.unc.edu}}

\affiliation{Department of Physics and Astronomy, University of North Carolina, 
Chapel Hill, NC 27599-3255, USA}

\date{\today}
\begin{abstract}
The minimal renormalizable $T^{'} \times Z_2$ model (MR$T^{'}$M) is slightly 
extended in its Higgs scalar sector such that the abelian part of the flavor 
symmetry enlarges to $(Z_2 \times Z_2^{'})$. All standard model and original 
MR$T^{'}$M states will transform trivially 
under $Z_2^{'}$. Inspired by the Valencia group's $A_4$ 
model building, we propose a $T^{'}$ WIMP candidate 
as the lightest $Z_2^{'}$ odd scalar. 
This extension of the prior MR$T^{'}$M model
maintains the successful predictions for the neutrino mixing matrix
and the Cabibbo angle, and provides an attractive candidate for dark matter ($\Phi_{WIMP}$) with
$M_{\Phi}\simeq 780$~GeV.
\end{abstract}


\maketitle

\section{Introduction}

\bigskip

\noindent
In recent theoretical cosmology, the most prominent issue is the dark side
of the universe consisting of two distinct sectors: dark matter
and dark energy. The latter sector is more difficult to explain
and may require a modification to the fundamental theory of gravity
or even redefining gravity
as an emergent property of the universe, like spacetime, rather than a 
fundamental force. Exciting new experimental data has led to a rapidly evolving
viewpoint on gravity over the past
few years and will hopefully lead to a better consensus.

\bigskip

Dark matter is much more approachable and is likely to be solved
more easily. It is, simply put, invisible matter which clumps like
luminous matter. Its existence has been known for 78 years,
since 1933~\cite{Zwicky}, while dark energy was discovered only 12 years ago
in 1998~\cite{Perlmutter,Riess}. The most popular candidate for  dark matter
is a Weakly Interacting Massive Particle (WIMP)~\cite{Goldberg}
with a mass range of $10^{2\pm1}$ GeV
and a typical weak interaction cross section with
standard model particles.

\bigskip

There are alternative dark matter candidates such as the invisible axion
with mass between $10^{-6}$~eV and $10^{-3}$~eV~\cite{DFS,Z,K,SVZ},
and Intermediate-Mass Black Holes (IMBHs)
with mass between $3 \times 10^2 M_{\odot}$ and
$3 \times 10^5 M_{\odot}$ ~\cite{FKTY,Carr}.
The fact that these candidate masses range over
77 orders of magnitude is indicative of the uncertainty
present in the problem. Unfortunately the well-known
galactic halo dark matter profiles
found from numerical simulations
are insensitive to the dark matter mass
because of scale invariance~\cite{NFW}.

\bigskip

Nevertheless, the WIMP is especially attractive because it naturally
gives the observed relic density. This is well known and will
be discussed later. The most popular candidate for a WIMP
was, at one time, the neutralino appearing in the supersymmetric
extension of the standard model~\cite{Goldberg}.  However, it has long been 
clear that a WIMP candidate does not require the assumption of
supersymmetry~\cite{Conf}. One non-supersymmetric
example of such a WIMP is the subject
of the present article.

\bigskip

All particles in the minimal standard model are badly suited for the role of
dark matter. Nevertheless,
the standard model has 28 free parameters when we include
massive neutrinos. Of these, no less than 22 arise from the masses and 
mixings of the quarks and leptons, 12 masses and 10 mixing angles. The
most popular approach towards explaining these 22 parameters is by
hypothesizing a flavor symmetry, $G_F$, which commutes with the
standard model gauge group. A promising choice for $G_F$ is
one of the finite non-abelian groups, $T^{'}$, the binary
tetrahedral flavor symmetry
\cite{FKtprime,FK2001,Feruglio:2007uu,Frampton:2007et,
Frampton:2008bz,CM,FKR,EFM1,EFM2,FHKM}.

\bigskip

The history of using $T^{'}$ as a flavor symmetry is lengthy and fascinating.
First used in Ref.~\cite{FKtprime} in 1994, it was implemented solely as a 
symmetry for quarks, because                                          
neutrinos were still believed to be massless. After neutrino masses and
mixings were discovered~\cite{SuperK}, the PMNS mixing matrix for leptons
was carefully measured and turned out to be very different from
the CKM mixing matrix for quarks. A number of different theories developed in
response to this first evidence of physics beyond the minimal standard model
~\cite{Frampton:1993wu,Aranda:2000tm,He:2003rm,Babu:2005se,Lee:2006pr}.
Eventually a useful approximation to the empirical PMNS matrix was 
determined to be the tribimaximal (TBM) one suggested in  Ref.~\cite{HPS}.

\bigskip

In the early 2000s, a purely leptonic flavor symmetry based on $A_4 = T$,
the tetrahedral group, was introduced by Ref.~\cite{Ma} to underpin
TBM mixing. Further investigation revealed that $A_4$ could not be extended 
to quarks because a viable CKM matrix could not be obtained~\cite{Altarelli}.
It was then realized that although $A_4$
is not a subgroup of its double cover~\cite{FKR}, $T^{'}$,
nevertheless from the viewpoint of the Kronecker products
used in particle theory  model building~\cite{FK2001}, $A_4$ behaves
{\it as if} it were a subgroup. This observation provided a
watershed where $T^{'}$ could act as a successful
flavor symmetry for quarks and leptons.

\section{Valencia mechanism}

\bigskip

\noindent
An ingenious new mechanism
has been discovered by a group based in Valencia~\cite{Valle1,Valle2}, 
working on $A_4$ model building.  Their implementation 
used the flavor symmetry group $A_4$, whose double cover is central to our
present work. 
It involved adding a small number of extra scalar fields, one of
which, by virtue of a discrete $Z_2$ analogous to R-symmetry in the MSSM, 
gives rise to stable dark matter. 

\bigskip

Their original model assigned all standard model leptons as different singlets
of $A_4$ with the right-handed neutrinos and one of the
newly added Higgs as the only $A_4$ triplets (the model's other Higgs was an 
$A_4$ singlet). These assignments were unconventional as most $A_4$
models, like the $T^{'}$ model discussed in later sections,
utilize triplets in the lepton assignments.

\bigskip

In Ref.~\cite{Valle1,Valle2} a particular
generator of $A_4$  was used to give rise to a $Z_2$ subgroup of $A_4$
that stabilized the WIMP.
This $Z_2$ group established a particle sector that is
discrete from the standard model particles and inaccessible
except via the weak force and gravity.

\bigskip

\noindent
Since $A_4$ alone has proved incapable of accommodating quarks in a like 
manner to leptons~\cite{Altarelli}, the Valencia group relegated the quark sector 
to future work. An alternative approach, that we pursue, is to use $T^{'}$ to 
replace $A_4$, allowing the incorporation of quarks, a prediction of the Cabibbo 
angle, and controllable deviations from TBM mixing angles.

\bigskip

\section{($T^{'} \times Z_2 \times Z_2^{'}$) Model}

\bigskip

\noindent
To accommodate the quark sector,
we adopt the ($T^{'} \times Z_2$) model formulated in
Ref.~\cite{Frampton:2008bz} and further analyzed in Ref.~\cite{EFM1, EFM2}.
This section will establish an extended model including elements of the Valencia
Mechanism by incorporating a second $Z_2^{'}$, while also
adding scalar fields and heavy right-handed neutrinos that are
odd under $Z_2{'}$; the lightest odd scalar will be the dark matter WIMP.
This model is a modification 
of the Minimal Renormalizable $T^{'}$ Model (MR$T^{'}$M) from
Ref.~\cite{EFM1} with a global symmetry of ($T^{'} \times Z_2 \times Z_2^{'}$)
restricting the Yukawa couplings.
One key difference from Ref.~\cite{Valle1,Valle2} is that 
$Z_2^{'}$ will not be subgroup of $T^{'}$.

\bigskip

The quark assignments below are unchanged from Ref.~\cite{Frampton:2008bz},
denoting ${\cal{Q}}_L=\dbinom{t}{b}_L$, $Q_L=\dbinom{c}{s}_L~\&~\dbinom{u}{d}_L$, ${\cal{C}}_R= c_R~\&~u_R$, and ${\cal{S}}_R= s_R~\&~d_R$. 
By setting all quarks even under $Z_2^{'}$,
past $T^{'}$ predictions are preserved.

\medskip
\begin{center}
\begin{tabular}{c||c|c|c|c|c|c|}
Quarks & ${\cal{Q}}_L$ & $Q_L$ & $t_R$ & $b_R$ & ${\cal{C}}_R$ & ${\cal{S}}_R$ 
\\ \hline \hline
$T^{'}$ & $1_1$ & $2_1$ & $1_1$ & $1_2$ & $2_3$ & $2_2$ \\ \hline
$Z_2$ & $+$ & $+$ & $+$ & $-$ & $-$ & $+$  \\ \hline
$Z_2^{'}$ & $+$ & $+$ & $+$ & $+$ & $+$ & $+$  \\ \hline
\end{tabular}
\end{center}
\medskip

\noindent
The lepton sector of Ref.~\cite{Frampton:2008bz} is retained unchanged,
even under $Z_2^{'}$, again keeping all the previous successes in 
Ref.~\cite{EFM1,EFM2}. Inspired by Ref.~\cite{Valle1,Valle2}, we have incorporated 
an additional triplet of right-handed neutrinos, $N_T$.  This triplet is odd under 
$Z_2^{'}$, and is summarized with the other lepton assignments below.

\medskip
\begin{center}
\begin{tabular}{c||c|c|c|c|c|c|c|c|}
Leptons & $L_L$ & $\tau_R$ & $\mu_R$ & $e_R$ & 
$N_R^{(1)}$ & $N_R^{(2)}$ & $N_R^{(3)}$ & $N_T$ \\ \hline \hline
$T^{'}$ & $3$ & $1_1$ & $1_2$ & $1_3$ & $1_1$ & $1_2$ & $1_3$ & $3$ \\ \hline
$Z_2$ & $+$ & $-$ & $-$ & $-$ & $+$ & $+$ & $+$ & $+$ \\ \hline
$Z_2^{'}$ & $+$ & $+$ & $+$ & $+$ & $+$ & $+$ & $+$ & $-$ \\ \hline
\end{tabular}
\end{center}
\medskip

\noindent
The Higgs sector is mostly the same as in Ref.~\cite{Frampton:2008bz}, 
$Z_2^{'}$-even, with a new $Z_2^{'}$-odd, $T^{'}$-triplet, $H_3^{''}$.
The five Higgs irreps of $T^{'}$ are shown in the following table.
Note that all of these scalars are doublets under the gauge group $SU(2)_L$.

\medskip
\begin{center}
\begin{tabular}{c||c|c|c|c|c|}
Higgs & $H_{1_1}$ & $H_{1_3}$ & $H_3$ & $H_3^{'}$ & $H_3^{''}$ \\ \hline \hline
$T^{'}$ & $1_1$ & $1_3$ & $3$ & $3$ & $3$ \\ \hline
$Z_2$ & $+$ & $-$ & $+$ & $-$ & $+$  \\ \hline
$Z_2^{'}$ & $+$ & $+$ & $+$ & $+$ & $-$ \\ \hline
\end{tabular}
\end{center}
\medskip

\bigskip

\noindent
The resultant Yukawa couplings are:

\bigskip

\begin{align}
{\cal{L}}_Y = M_0 N_T N_T +  M_1 N_R^{(1)} N_R^{(1)} +
 M_{23} N_R^{(2)} N_R^{(3)} + \nonumber \\
Y_{e} L_L e_R H_3^{'} + Y_{\mu} L_L \mu_R H_3^{'} + 
Y_{\tau} L_L \tau_R H_3^{'} + \nonumber \\
Y_1 L_L N_R^{(1)} H_3 + Y_2 L_L N_R^{(2)} H_3 + 
Y_3 L_L N_R^{(3)} H_3 + \nonumber \\
Y_4 L_L (N_T H_3^{''})_3 + Y_5 L_L (N_T H_3^{''})_{3^{'}} + \nonumber \\
Y_t ({\cal{Q}}_L t_R H_{1_1})  + Y_b ({\cal{Q}}_L b_R H_{1_3}) + \nonumber \\
Y_{\cal C} (Q_L {\cal{C}}_R H^{'}_{3}) + Y_{\cal S} (Q_L {\cal{S}}_R H_{3})  + h.c. \ .
\end{align}

\bigskip

It is interesting to note that the terms with the 
new right-handed neutrino triplet, $N_T$,
and new Higgs, $H_3^{''}$, involve 
($3 \times 3 \times 3$) under $T^{'}$, which contains
two ($1_1$) singlets~\cite{TW}, and hence produces just 
two additional Yukawa couplings.  This will prove 
important to our implementation of the Type-I seesaw mechanism.
The same Yukawa couplings, 
with $Y_4$ and/or $Y_5$ complexified,
will naturally lead to leptogenesis~\footnote{It is notable that one decay mode 
of the triplet $N_T$ is into a light neutrino and dark matter.}.

\bigskip

\section{Dark Matter and Neutrino Predictions}

\bigskip

\subsection{Dark Matter Candidate}
\noindent
The $T^{'}$ WIMP candidate is the lightest state with an assignment 
of $Z_2^{'} = -1$. The 
$Z_2^{'}$ odd states are $N_T$ and $H_3^{''}$. The neutrino triplet, $N_T$, is 
expected to be very heavy from the seesaw mechanism discussed in the 
Appendix~\ref{sec:seesaw}. It decays into an $H_3^{''}$ and a 
lepton, making it a good candidate for the leptogenesis mechanism~\cite{FY}. 

\bigskip

The WIMP candidate is therefore a superposition 
of the CP-even neutral scalars contained in $H^{''}_3$, which
has three $SU(2)_L$ doublets:
\begin{equation}
H_3^{''}(1)=
\begin{pmatrix}
h_1^+  \\
h_1^0 + i A_1 
\end{pmatrix} 
,~H_3^{''}(2)=
\begin{pmatrix}
h_2^+  \\
h_2^0 + i A_2 
\end{pmatrix}
,~H_3^{''}(3)=
\begin{pmatrix}
h_3^+  \\
h_3^0 + i A_3 
\end{pmatrix}~.
\end{equation}
This set includes 6 charged scalars, 3 neutral CP-even scalars, and 3 neutral 
CP-odd scalars. Our dark matter candidate will be a superposition of the three real
$Z_2^{'}$-odd, CP-even, neutral scalar states:
\begin{equation}
\Phi_{WIMP} =\alpha h_{1}^{0}+\beta h_{2}^{0}+\gamma h_{3}^{0}~.
\label{WIMPbits}
\end{equation}
An evaluation of the dark matter candidate coefficients, 
$\alpha$, $\beta$, and $\gamma$, requires
knowledge of the coefficients
in the Higgs scalar potential, shown in Appendix~\ref{sec:portent}, 
and is beyond the scope of this paper.

\bigskip

\subsection{Relic Density and WIMP Mass}
\noindent
A common tool for estimating the mass of a dark matter candidate ($M_{\Phi}$) is the relic 
density. This approach uses both particle and cosmological inputs as well as model 
properties to estimate the annihilation cross section and particle density after 
freeze out. We will follow the general treatment outlined in Ref.~\cite{strumia}. 
Starting with,
\begin{equation}
\frac{n_{DM} (T)}{s(T)}\approx\biggl(\frac{1}{M_P T_f 
\langle\sigma_A v\rangle}\biggr)\sqrt{\frac{180}{\pi g_C}},
\end{equation}
and noting a weak hypercharge of $Y=1/2$ and the Planck mass of 
$M_P=1.22\times10^{19}$~GeV. Also, in the relevant 
temperature range ($T_f\ll M_{\Phi}$), we can safely approximate 
$M_{\Phi}/T_f\approx26$.  Focusing on an
approximation of the most significant instance of elastic scattering, we will adopt the 
equation seen in
Ref.~\cite{Valle1},
\begin{equation}
\langle\sigma_A v\rangle\simeq\frac{3g_2^4 + g_Y^4 + 6g_2^2g_Y^2 + 
4\lambda^2}{128\pi M_{\Phi}^2}
\end{equation}
where $g_2=\sqrt{(4\pi\alpha)/(1-(M_W/W_Z)^2)}$ and 
$g_Y=\sqrt{4\pi\alpha}M_Z/M_W$. Rather than solve the Higgs 
scalar potential (detailed in Appendix~\ref{sec:portent}), we make the assumption 
that the quartic coupling constant, $\lambda$, yields a very small contribution. This allows us 
to simplify the relic density to the form,
\begin{equation}
\frac{n_{DM} (T)}{s(T)}\approx\biggl(\frac{1248 M_{\Phi}M_W^4(M_Z^2-M_W^2)^2}{\alpha^2M_P M_Z^4 (M_Z^4+4M_Z^2M_W^2-2M_W^4)}\biggr)\sqrt{\frac{5}{\pi^3 g_C}}.
\end{equation}
Data from WMAP 7 indicates $\Omega_{DM}h^2=0.110\pm0.006$ (i.e.
 $n_{DM}/s=0.40\pm0.02$~eV/$M_{\Phi}$)~\cite{WMAP} and from
Ref.~\cite{pdg} indicates $M_W=80.399$~GeV and $M_Z=91.1876$~GeV. Finally 
we note that $g_C$, the degrees of freedom, is based (at this temperature range) on 
$g=(g_B+\frac{7}{8}g_F)$ from the numbers of bosons and fermions. 
The degrees of freedom can also be split into $g_{SM}=104.125$ for the standard 
model (using Majorana neutrinos unlike the more commonly referenced $106.75$ 
assuming Dirac neutrinos) and $1\lesssim n\lesssim 37.25$ for our model's additions. 
Taking the maximally allowed degrees of freedom results in $g_C\approx141.375$ and 
$M_{\Phi}\approx0.78$~TeV.

\bigskip

\subsection{Neutrino Mixing}
\noindent
By relying on an exterior $Z_2^{'}$ group rather 
than a subgroup of $T^{'}$, the quark assignments and couplings
have been left identical to that of previous $T^{'}$ models.  
This serves to preserve the predictions of the Cabibbo angle
~\cite{Frampton:2008bz},
\begin{equation}
\tan{2 \Theta_{12}}=\frac{\sqrt{2}}{3}.
\end{equation}
Also preserved is the MR$T^{'}$M seesaw mechanism relation (detailed in 
Appendix~\ref{sec:seesaw}) between reactor and 
atmospheric neutrino mixing angles, made possible by perturbing the Cabibbo angle 
closer to its empirical value~\cite{EFM1}.
\begin{equation}
\theta_{13}=\sqrt{2}~(\frac{\pi}{4}-\theta_{23}).
\label{eqn:EFM}
\end{equation}
Current best fit estimates of the neutrino 
mixing angles $\theta_{13}$ and  $\theta_{23}$, are based on a measurement of 
$2\sin^2{\theta_{23}}\sin^2{2\theta_{13}}$ (commonly listed as $\sin^2{2\theta_{13}}$ 
under the assumption of maximal $\theta_{23}$). By combining Eq.~(\ref{eqn:EFM}) with the
recent integration of experimental data from T2K+MINOS+DC~\cite{theta13}, 
we can solve for unique sets of these two neutrino angles. Based on measurements of 
$0.015 < 2\sin^2{\theta_{23}}\sin^2{2\theta_{13}} < 0.15$ at $95\%$~CL with a best fit value 
of $0.08$ (assuming normal hierarchy), we predict the following values,
\begin{equation}
\theta_{23} =  38_{~~-3^{\circ}}^{\circ~+4^{\circ}} , ~~~~~~~~~~  \theta_{13} =   9_{~~-5^{\circ}}^{\circ~+5^{\circ}}.
\label{eqn:1323}
\end{equation}
The values in Eq.~(\ref{eqn:1323}) are significant deviations from TBM values and 
should soon be verified by neutrino experiments.

\bigskip

\section{Discussion}

\bigskip

\noindent
The use of $T^{'}$ flavor symmetry has previously
led to interesting predictions for the CKM and PMNS
mixing matrices for quarks and leptons, respectively.
Of special interest is the fact that $T^{'}$
flavor symmetry predicts a unique
relationship between these two mixing
matrices that enter the $W^{\pm}$ mixings of
weak interactions. More accurate experimental
measurements of the mixing angles, particularly
$\theta_{13}$, are presently underway, and it will
be interesting to discover
whether the $T^{'}$ predictions are corroborated. 

\bigskip

There is wide expectation that the LHC will
shortly, perhaps within a year, discover the Higgs 
boson (H). Of special significance to the present
model are the H production cross section and
the H partial decay widths. These depend
theoretically on the Yukawa coupling of H
to the fermions, the quarks, and leptons. In the minimal standard model (MSM)
these Yukawa couplings are all simply proportional
to the fermion mass, implying that the H production by
gluon fusion is dominated by a one-loop top quark triangle
with dominant decay modes of bottom quarks
and tau leptons. In the $T^{'}$ model, the Yukawa
couplings do not follow this pattern,
and significant deviations from the minimal
standard model are expected. Although everything
else about the MSM has withstood close scrutiny,
the Yukawa couplings are not geometrical like the
gauge couplings, and appear as the most
vulnerable piece of the MSM Lagrangian.

\bigskip

In this article, we have shown how the $T^{'}$ model
can be adapted to include a WIMP dark matter candidate
without modifying any prior predictions. Hopefully, new data from the LHC and 
assorted neutrino experiments will soon allow us to confirm this $T^{'}$ model.

\newpage

\appendix
\section{Generalized Type-I Seesaw Mechanism}
\label{sec:seesaw}

\noindent
At this point we can state that the vacuum expectation values (VEVs) of our model's Higgs are as follows,

\begin{equation}
\begin{array}{c}
<H_3>~=~(V_1,V_2,V_3) ,\\
<H_3^{'}>~=~(\frac{m_{\tau}}{Y_{\tau}},\frac{m_{\mu}}{Y_{\mu}},\frac{m_{e}}{Y_{e}}) ,\\
<H_3^{''}>~=~(0,0,0) ,\\
<H_{1_{1}}>=(\frac{m_t}{Y_t}),~<H_{1_{3}}>=(\frac{m_b}{Y_b})~.
\label{VEVs}
\end{array}
\end{equation}

\noindent
$H_3^{'}$ is tied to the charged lepton masses and remains disconnected from the neutrinos 
assuming the charged leptons are mass eigenstates. $H_3^{''}$ must have at least one component 
without a VEV in order to create stable dark matter but must also have 3 identical values in order 
for $Z_2^{'}$ to commute with $T^{'} \times Z_2$, hence three zeroes. $H_3$ remains in 
a general form that we will further specify by using the seesaw mechanism. 

\bigskip

We will begin with the tribimaximal form of the PMNS mixing matrix, an analog of the CKM matrix for 
neutrinos.  
\begin{equation}
\begin{pmatrix}
\nu_1 \\
\nu_2 \\
\nu_3
\end{pmatrix}
=
U_{TBM}
\begin{pmatrix}
\nu_{e} \\
\nu_{\mu} \\
\nu_{\tau}
\end{pmatrix}
,
U_{TBM}=
\begin{pmatrix}
\sqrt{\frac{2}{3}} & \sqrt{\frac{1}{3}} & 0 \\
-\sqrt{\frac{1}{6}} & \sqrt{\frac{1}{3}} & -\sqrt{\frac{1}{2}} \\
-\sqrt{\frac{1}{6}} & \sqrt{\frac{1}{3}} & \sqrt{\frac{1}{2}}
\end{pmatrix}
\end{equation}
\noindent
The PMNS matrix can be used to diagonalize the neutrino mass matrix, and by using the 
tribimaximal form we can predict the preferred symmetry.
\begin{equation}
\begin{array}{c}
 M_{\nu,diag}=U_{TBM}^{T} M_{\nu} U_{TBM} ,\\  \\
 M_{\nu}=U_{TBM}
\begin{pmatrix}
m_1 & 0 & 0 \\
0 & m_2 & 0 \\
0 & 0 & m_3
\end{pmatrix}
U_{TBM}^{T}
\end{array}
\end{equation}
\noindent
The resultant matrix, presented here in simplified form, 
will then be compared with the same matrix 
derived via other means.
\begin{equation}
M_{\nu}=
\begin{pmatrix}
-A+B+C & A & A \\
A & B & C \\
A & C & B
\end{pmatrix}
\label{sym}
\end{equation}

\bigskip

Next we will implement a generalized Type-I Seesaw Mechanism (the $(3,6)$ form defined 
by 3 families and 6 $SU(2)$ singlet field)~\cite{SV}, first noting the key equation in 
Ref.~\cite{Mink} showing another way to determine $M_{\nu}$,
\begin{equation}
M_{\nu}=M_D M_N^{-1} M_D^T .
\label{eqn:neumass}
\end{equation}

\bigskip

The Dirac and Majorana mass matrices below are based on a generalized form of those 
used in Ref.~\cite{Frampton:2008ci}. Due to the 6 right-handed neutrino states, the Majorana 
matrix enlarges to $6\times6$, while the Dirac matrix becomes $3\times6$. The zero 
elements of the Dirac mass matrix are caused by VEV zeroes of $H_3^{''}$.
\begin{equation}
M_D=
\begin{pmatrix}
0 & 0 & 0 & Y_1 V_1 & Y_2 V_3 & Y_3 V_2 \\
0 & 0 & 0 & Y_1 V_3 & Y_2 V_2 & Y_3 V_1 \\
0 & 0 & 0 & Y_1 V_2 & Y_2 V_1 & Y_3 V_3 
\end{pmatrix} 
,~M_N=
\begin{pmatrix}
M_0 & 0 & 0 & 0 & 0 & 0 \\
0 & M_0 & 0 & 0 & 0 & 0 \\
0 & 0 & M_0 & 0 & 0 & 0 \\
0 & 0 & 0 & M_1 & 0 & 0 \\
0 & 0 & 0 & 0 & 0 & M_{23} \\
0 & 0 & 0 & 0 & M_{23} & 0 
\end{pmatrix}
\end{equation}
For simplicity we will set 
$x_1 \equiv Y_1^2/M_1$ and $x_{23} \equiv Y_2 Y_3/M_{23}$.   By following 
Eq.~(\ref{eqn:neumass}) we find the symmetric form of the light neutrino mass matrix,
\begin{equation}
M_{\nu}=
\begin{pmatrix}
x_1 V_1^2 + 2 x_{23} V_2 V_3 & & x_1 V_1 V_3 + X_{23} (V_2^2 + V_1 V_3) & & x_1 V_1 V_2 + x_{23} (V_3^2 + V_1 V_2) \\
 & & x_1 V_3^2 + 2 x_{23} V_1 V_2 & & x_1 V_2 V_3 + x_{23} (V_1^2 + V_2 V_3) \\
 & & & & x_1 V_2^2 + 2 x_{23} V_1 V_3
\end{pmatrix}
\label{M_mu}
\end{equation}
After comparing Eq.~(\ref{M_mu}) to its symmetric counter part, Eq.~(\ref{sym}), we attempt to 
solve for the VEVs of $H_3$.  One possibility that preserves an 
acceptable form of the neutrino masses is~$<H_3>=(-2,1,1)$.  These values can then be 
plugged into the eigenvalues of Eq.~(\ref{sym}) resulting in the parameterized values of the 
left-handed neutrino masses.
\begin{equation}
\begin{array}{l}
m_1=B+C-2A=-9 x_{23} \\
m_2=A+B+C=0 \\
m_3=B-C=6 x_1 + 3 x_{23}
\end{array}
\end{equation}

These solutions show that the addition of the neutrino triplet to the MR$T^{'}$M does not 
change the results of the seesaw mechanism and preserves the predictions of 
Ref.~\cite{Frampton:2008bz,EFM1,EFM2}.

\newpage

\section{The Higgs Scalar Potential}
\label{sec:portent}
\noindent
Included below is the Higgs scalar potential up to quartic order, consisting of 218 terms 
including 77 hermitian conjugates. We will use $1_{1,2,3}$, 
to represent the three singlet representations of $T^{'}$; additionally $3_1$ and 
$3_2$ will be used to distinguish the two triplet products of two contracted $T^{'}$ triplets.

We have studied assiduously the set of 
equations $\partial V/ \partial v_i$, where the $v_i$ are the VEVs, and the related requirements for a local minimum of positive
Hessian eigenvalues. We find, after careful calculation,
that the VEVs in Eq.~(\ref{VEVs}) are allowed without fine tuning.

Without further assumptions, one cannot determine the superposition coefficients 
$\alpha$, $\beta$, and $\gamma$ of 
Eq.~(\ref{WIMPbits}). It may be fruitful to seek
an additional assumption to increase our model's
predictivity. For the dedicated reader who wishes
to pursue this interesting question, we provide below
the complete Higgs potential.
\small
\begin{gather*}
V=\mu_{H_{1_{1}}}^2 H_{1_1}^{\dagger} H_{1_1}+\mu_{H_{1_{3}}}^2 H_{1_3}^{\dagger} H_{1_3}+
\mu_{H_{3}}^2 H_{3}^{\dagger} H_{3}+\mu_{H_{3}^{'}}^2 H_{3}^{' \dagger} H_{3}^{'}+\mu_{H_{3}^{''}}^2 H_{3}^{'' \dagger} H_{3}^{''}  +
\lambda_{1} [H_{1_{1}}^{\dagger} H_{1_{1}}]_{1_{1}}^2 \\+
\lambda_{2} [H_{1_{3}}^{\dagger} H_{1_{3}}]_{1_{1}}^2+
\lambda_{3} [H_{1_{1}}^{\dagger} H_{1_{1}}]_{1_{1}} [H_{1_{3}}^{\dagger} H_{1_{3}}]_{1_{1}}+
\lambda_{4} [H_{1_{1}}^{\dagger} H_{1_{3}}^{\dagger}]_{1_{2}} [H_{1_{1}} H_{1_{3}}]_{1_{3}}+
\lambda_{5} [H_{1_{3}}^{\dagger} H_{1_{3}}^{\dagger}]_{1_{3}} [H_{1_{3}} H_{1_{3}}]_{1_{2}} \\+
\lambda_{6} [H_{1_{1}}^{\dagger} H_{1_{1}}]_{1_{1}} [H_{3}^{\dagger} H_{3}]_{1_{1}} +
\lambda_{7} [H_{1_{1}}^{\dagger} H_{1_{1}}]_{1_{1}} [H_{3}^{' \dagger} H_{3}^{'}]_{1_{1}} +
\lambda_{8} [H_{1_{1}}^{\dagger} H_{1_{1}}]_{1_{1}} [H_{3}^{'' \dagger} H_{3}^{''}]_{1_{1}} \\+
\lambda_{9} [H_{1_{3}}^{\dagger} H_{1_{3}}]_{1_{1}} [H_{3}^{\dagger} H_{3}]_{1_{1}} +
\lambda_{10} [H_{1_{3}}^{\dagger} H_{1_{3}}]_{1_{1}} [H_{3}^{' \dagger} H_{3}^{'}]_{1_{1}} +
\lambda_{11} [H_{1_{3}}^{\dagger} H_{1_{3}}]_{1_{1}} [H_{3}^{'' \dagger} H_{3}^{''}]_{1_{1}} \\+
\lambda_{12} ([H_{1_{1}}^{\dagger} H_{1_{1}}^{\dagger}]_{1_{1}} [H_{3} H_{3}]_{1_{1}} +h.c.)+
\lambda_{13} ([H_{1_{1}}^{\dagger} H_{1_{1}}^{\dagger}]_{1_{1}} [H_{3}^{'} H_{3}^{'}]_{1_{1}} +h.c.)+
\lambda_{14} ([H_{1_{1}}^{\dagger} H_{1_{1}}^{\dagger}]_{1_{1}} [H_{3}^{''} H_{3}^{''}]_{1_{1}} +h.c.) \\+
\lambda_{15} ([H_{1_{3}}^{\dagger} H_{1_{3}}^{\dagger}]_{1_{3}} [H_{3} H_{3}]_{1_{2}} +h.c.)+
\lambda_{16} ([H_{1_{3}}^{\dagger} H_{1_{3}}^{\dagger}]_{1_{3}} [H_{3}^{'} H_{3}^{'}]_{1_{2}} +h.c.)+
\lambda_{17} ([H_{1_{3}}^{\dagger} H_{1_{3}}^{\dagger}]_{1_{3}} [H_{3}^{''} H_{3}^{''}]_{1_{2}} +h.c.)\\+
\lambda_{18} [H_{1_{1}}^{\dagger} H_{3}]_{3} [H_{3}^{\dagger} H_{1_{1}}]_{3} +
\lambda_{19} [H_{1_{1}}^{\dagger} H_{3}^{'}]_{3} [H_{3}^{' \dagger} H_{1_{1}}]_{3} +
\lambda_{20} [H_{1_{1}}^{\dagger} H_{3}^{''}]_{3} [H_{3}^{'' \dagger} H_{1_{1}}]_{3} \\+
\lambda_{21} [H_{1_{3}}^{\dagger} H_{3}]_{3} [H_{3}^{\dagger} H_{1_{3}}]_{3} +
\lambda_{22} [H_{1_{3}}^{\dagger} H_{3}^{'}]_{3} [H_{3}^{' \dagger} H_{1_{3}}]_{3} +
\lambda_{23} [H_{1_{3}}^{\dagger} H_{3}^{''}]_{3} [H_{3}^{'' \dagger} H_{1_{3}}]_{3} \\+
\lambda_{24} ([H_{1_{3}}^{\dagger} H_{3}^{'}]_{3} [H_{3}^{\dagger} H_{1_{1}}]_{3} +h.c.) \\+
\lambda_{25} ([H_{1_{1}}^{\dagger} H_{3}]_{3} [H_{3}^{\dagger} H_{3}]_{3_{1}} +h.c.) +
\lambda_{26} ([H_{1_{1}}^{\dagger} H_{3}^{\dagger}]_{3} [H_{3} H_{3}]_{3_{1}} +h.c.) \\+
\lambda_{27} ([H_{1_{1}}^{\dagger} H_{3}]_{3} [H_{3}^{\dagger} H_{3}]_{3_{2}} +h.c.) +
\lambda_{28} ([H_{1_{1}}^{\dagger} H_{3}^{\dagger}]_{3} [H_{3} H_{3}]_{3_{2}} +h.c.) \\+
\lambda_{29} ([H_{1_{1}}^{\dagger} H_{3}]_{3} [H_{3}^{' \dagger} H_{3}^{'}]_{3_{1}} +h.c.)+
\lambda_{30} ([H_{1_{1}}^{\dagger} H_{3}^{'}]_{3} [H_{3}^{\dagger} H_{3}^{'}]_{3_{1}} +h.c.)\\+
\lambda_{31} ([H_{1_{1}}^{\dagger} H_{3}^{\dagger}]_{3} [H_{3}^{'} H_{3}^{'}]_{3_{1}} +h.c.)+
\lambda_{32} ([H_{1_{1}}^{\dagger} H_{3}^{' \dagger}]_{3} [H_{3} H_{3}^{'}]_{3_{1}} +h.c.)\\+
\lambda_{33} ([H_{1_{1}}^{\dagger} H_{3}]_{3} [H_{3}^{' \dagger} H_{3}^{'}]_{3_{2}} +h.c.)+
\lambda_{34} ([H_{1_{1}}^{\dagger} H_{3}^{'}]_{3} [H_{3}^{\dagger} H_{3}^{'}]_{3_{2}} +h.c.)\\+
\lambda_{35} ([H_{1_{1}}^{\dagger} H_{3}^{\dagger}]_{3} [H_{3}^{'} H_{3}^{'}]_{3_{2}} +h.c.)+
\lambda_{36} ([H_{1_{1}}^{\dagger} H_{3}^{' \dagger}]_{3} [H_{3} H_{3}^{'}]_{3_{2}} +h.c.)\displaybreak\\ \\ \\+
\lambda_{37} ([H_{1_{1}}^{\dagger} H_{3}]_{3} [H_{3}^{'' \dagger} H_{3}^{''}]_{3_{1}} +h.c.)+
\lambda_{38} ([H_{1_{1}}^{\dagger} H_{3}^{''}]_{3} [H_{3}^{\dagger} H_{3}^{''}]_{3_{1}} +h.c.)\\+
\lambda_{39} ([H_{1_{1}}^{\dagger} H_{3}^{\dagger}]_{3} [H_{3}^{''} H_{3}^{''}]_{3_{1}} +h.c.)+
\lambda_{40} ([H_{1_{1}}^{\dagger} H_{3}^{'' \dagger}]_{3} [H_{3} H_{3}^{''}]_{3_{1}} +h.c.)\\+
\lambda_{41} ([H_{1_{1}}^{\dagger} H_{3}]_{3} [H_{3}^{'' \dagger} H_{3}^{''}]_{3_{2}} +h.c.)+
\lambda_{42} ([H_{1_{1}}^{\dagger} H_{3}^{''}]_{3} [H_{3}^{\dagger} H_{3}^{''}]_{3_{2}} +h.c.)\\+
\lambda_{43} ([H_{1_{1}}^{\dagger} H_{3}^{\dagger}]_{3} [H_{3}^{''} H_{3}^{''}]_{3_{2}} +h.c.)+
\lambda_{44} ([H_{1_{1}}^{\dagger} H_{3}^{'' \dagger}]_{3} [H_{3} H_{3}^{''}]_{3_{2}} +h.c.) \\+
\lambda_{45} ([H_{1_{3}}^{\dagger} H_{3}^{'}]_{3} [H_{3}^{' \dagger} H_{3}^{'}]_{3_{1}} +h.c.) +
\lambda_{46} ([H_{1_{3}}^{\dagger} H_{3}^{' \dagger}]_{3} [H_{3}^{'} H_{3}^{'}]_{3_{1}} +h.c.) \\+
\lambda_{47} ([H_{1_{3}}^{\dagger} H_{3}^{'}]_{3} [H_{3}^{' \dagger} H_{3}^{'}]_{3_{2}} +h.c.) +
\lambda_{48} ([H_{1_{3}}^{\dagger} H_{3}^{' \dagger}]_{3} [H_{3}^{'} H_{3}^{'}]_{3_{2}} +h.c.) \\+
\lambda_{49} ([H_{1_{3}}^{\dagger} H_{3}^{'}]_{3} [H_{3}^{\dagger} H_{3}]_{3_{1}} +h.c.)+
\lambda_{50} ([H_{1_{3}}^{\dagger} H_{3}]_{3} [H_{3}^{' \dagger} H_{3}]_{3_{1}} +h.c.)\\+
\lambda_{51} ([H_{1_{3}}^{\dagger} H_{3}^{' \dagger}]_{3} [H_{3} H_{3}]_{3_{1}} +h.c.) +
\lambda_{52} ([H_{1_{3}}^{\dagger} H_{3}^{\dagger}]_{3} [H_{3}^{'} H_{3}]_{3_{1}} +h.c.) \\+
\lambda_{53} ([H_{1_{3}}^{\dagger} H_{3}^{'}]_{3} [H_{3}^{\dagger} H_{3}]_{3_{2}} +h.c.) +
\lambda_{54} ([H_{1_{3}}^{\dagger} H_{3}]_{3} [H_{3}^{' \dagger} H_{3}]_{3_{2}} +h.c.) \\+
\lambda_{55} ([H_{1_{3}}^{\dagger} H_{3}^{' \dagger}]_{3} [H_{3} H_{3}]_{3_{2}} +h.c.) +
\lambda_{56} ([H_{1_{3}}^{\dagger} H_{3}^{\dagger}]_{3} [H_{3}^{'} H_{3}]_{3_{2}} +h.c.) \\+
\lambda_{57} ([H_{1_{3}}^{\dagger} H_{3}^{'}]_{3} [H_{3}^{'' \dagger} H_{3}^{''}]_{3_{1}} +h.c.) +
\lambda_{58} ([H_{1_{3}}^{\dagger} H_{3}^{''}]_{3} [H_{3}^{' \dagger} H_{3}^{''}]_{3_{1}} +h.c.) \\+
\lambda_{59} ([H_{1_{3}}^{\dagger} H_{3}^{' \dagger}]_{3} [H_{3}^{''} H_{3}^{''}]_{3_{1}} +h.c.) +
\lambda_{60} ([H_{1_{3}}^{\dagger} H_{3}^{'' \dagger}]_{3} [H_{3}^{'} H_{3}^{''}]_{3_{1}} +h.c.) \\+
\lambda_{61} ([H_{1_{3}}^{\dagger} H_{3}^{'}]_{3} [H_{3}^{'' \dagger} H_{3}^{''}]_{3_{2}} +h.c.) +
\lambda_{62} ([H_{1_{3}}^{\dagger} H_{3}^{''}]_{3} [H_{3}^{' \dagger} H_{3}^{''}]_{3_{2}} +h.c.) \\+
\lambda_{63} ([H_{1_{3}}^{\dagger} H_{3}^{' \dagger}]_{3} [H_{3}^{''} H_{3}^{''}]_{3_{2}} +h.c.) +
\lambda_{64} ([H_{1_{3}}^{\dagger} H_{3}^{'' \dagger}]_{3} [H_{3}^{'} H_{3}^{''}]_{3_{2}} +h.c.) \\+
\lambda_{65} [H_{3}^{\dagger} H_{3}]_{1_{2}} [H_{3}^{\dagger} H_{3}]_{1_{3}} +
\lambda_{66} [H_{3}^{\dagger} H_{3}^{\dagger}]_{1_{2}} [H_{3} H_{3}]_{1_{3}} +
\lambda_{66}^{'} [H_{3}^{\dagger} H_{3}^{\dagger}]_{1_{3}} [H_{3} H_{3}]_{1_{2}} \\+
\lambda_{67} [H_{3}^{\dagger} H_{3}]_{1_1}^2 +
\lambda_{68} [H_{3}^{\dagger} H_{3}^{\dagger}]_{1_{1}} [H_{3} H_{3}]_{1_{1}} +
\lambda_{69} ([H_{3}^{\dagger} H_{3}]_{3_{1}} [H_{3}^{\dagger} H_{3}]_{3_{1}} +h.c.) \\+
\lambda_{70} [H_{3}^{\dagger} H_{3}]_{3_{1}} [H_{3}^{\dagger} H_{3}]_{3_{2}} +
\lambda_{71} [H_{3}^{\dagger} H_{3}^{\dagger}]_{3_{1}} [H_{3} H_{3}]_{3_{2}}  \\+
\lambda_{72} [H_{3}^{' \dagger} H_{3}^{'}]_{1_{2}} [H_{3}^{' \dagger} H_{3}^{'}]_{1_{3}} +
\lambda_{73} [H_{3}^{' \dagger} H_{3}^{' \dagger}]_{1_{2}} [H_{3}^{'} H_{3}^{'}]_{1_{3}} +
\lambda_{73}^{'} [H_{3}^{' \dagger} H_{3}^{' \dagger}]_{1_{3}} [H_{3}^{'} H_{3}^{'}]_{1_{2}} \\+
\lambda_{74} [H_{3}^{' \dagger} H_{3}^{'}]_{1_1}^2 +
\lambda_{75} [H_{3}^{' \dagger} H_{3}^{' \dagger}]_{1_{1}} [H_{3}^{'} H_{3}^{'}]_{1_{1}} +
\lambda_{76} ([H_{3}^{' \dagger} H_{3}^{'}]_{3_{1}} [H_{3}^{' \dagger} H_{3}^{'}]_{3_{1}} +h.c.) \\+
\lambda_{77} [H_{3}^{' \dagger} H_{3}^{'}]_{3_{1}} [H_{3}^{' \dagger} H_{3}^{'}]_{3_{2}} +
\lambda_{78} [H_{3}^{' \dagger} H_{3}^{' \dagger}]_{3_{1}} [H_{3}^{'} H_{3}^{'}]_{3_{2}}  \\+
\lambda_{79} [H_{3}^{'' \dagger} H_{3}^{''}]_{1_{2}} [H_{3}^{'' \dagger} H_{3}^{''}]_{1_{3}} +
\lambda_{80} [H_{3}^{'' \dagger} H_{3}^{'' \dagger}]_{1_{2}} [H_{3}^{''} H_{3}^{''}]_{1_{3}} +
\lambda_{80}^{'} [H_{3}^{'' \dagger} H_{3}^{'' \dagger}]_{1_{3}} [H_{3}^{''} H_{3}^{''}]_{1_{2}} \\+
\lambda_{81} [H_{3}^{'' \dagger} H_{3}^{''}]_{1_1}^2 +
\lambda_{82} [H_{3}^{'' \dagger} H_{3}^{'' \dagger}]_{1_{1}} [H_{3}^{''} H_{3}^{''}]_{1_{1}} +
\lambda_{83} ([H_{3}^{'' \dagger} H_{3}^{''}]_{3_{1}} [H_{3}^{'' \dagger} H_{3}^{''}]_{3_{1}} +h.c.) \\+
\lambda_{84} [H_{3}^{'' \dagger} H_{3}^{''}]_{3_{1}} [H_{3}^{'' \dagger} H_{3}^{''}]_{3_{2}} +
\lambda_{85} [H_{3}^{'' \dagger} H_{3}^{'' \dagger}]_{3_{1}} [H_{3}^{''} H_{3}^{''}]_{3_{2}}\displaybreak \\ \\+
\lambda_{86} [H_{3}^{\dagger} H_{3}]_{1_{1}} [H_{3}^{' \dagger} H_{3}^{'}]_{1_{1}}+
\lambda_{87} ([H_{3}^{\dagger} H_{3}^{'}]_{1_{1}} [H_{3}^{\dagger} H_{3}^{'}]_{1_{1}} +h.c.)\\+
\lambda_{88} [H_{3}^{\dagger} H_{3}^{' \dagger}]_{1_{1}} [H_{3} H_{3}^{'}]_{1_{1}} +
\lambda_{89} ([H_{3}^{\dagger} H_{3}^{\dagger}]_{1_{1}} [H_{3}^{'} H_{3}^{'}]_{1_{1}} +h.c.)\\+
\lambda_{90} [H_{3}^{\dagger} H_{3}]_{1_{2}} [H_{3}^{' \dagger} H_{3}^{'}]_{1_{3}} +
\lambda_{91} ([H_{3}^{\dagger} H_{3}^{'}]_{1_{2}} [H_{3}^{\dagger} H_{3}^{'}]_{1_{3}} +h.c.)\\+
\lambda_{92} [H_{3}^{\dagger} H_{3}^{' \dagger}]_{1_{2}} [H_{3} H_{3}^{'}]_{1_{3}} +
\lambda_{93} ([H_{3}^{\dagger} H_{3}^{\dagger}]_{1_{2}} [H_{3}^{'} H_{3}^{'}]_{1_{3}} +h.c.)\\+
\lambda_{92}^{'} [H_{3}^{\dagger} H_{3}^{' \dagger}]_{1_{3}} [H_{3} H_{3}^{'}]_{1_{2}} +
\lambda_{93}^{'} ([H_{3}^{\dagger} H_{3}^{\dagger}]_{1_{3}} [H_{3}^{'} H_{3}^{'}]_{1_{2}} +h.c.)\\+
\lambda_{94} ([H_{3}^{\dagger} H_{3}]_{3_{1}} [H_{3}^{' \dagger} H_{3}^{'}]_{3_{1}} +h.c.)+
\lambda_{95} ([H_{3}^{\dagger} H_{3}^{'}]_{3_{1}} [H_{3}^{\dagger} H_{3}^{'}]_{3_{1}} +h.c.)\\+
\lambda_{96} [H_{3}^{\dagger} H_{3}]_{3_{1}} [H_{3}^{' \dagger} H_{3}^{'}]_{3_{2}} +
\lambda_{97} ([H_{3}^{\dagger} H_{3}^{'}]_{3_{1}} [H_{3}^{\dagger} H_{3}^{'}]_{3_{2}} +h.c.)\\+
\lambda_{98} [H_{3}^{\dagger} H_{3}^{' \dagger}]_{3_{1}} [H_{3} H_{3}^{'}]_{3_{2}} +
\lambda_{99} ([H_{3}^{\dagger} H_{3}^{\dagger}]_{3_{1}} [H_{3}^{'} H_{3}^{'}]_{3_{2}} +h.c.)\\+
\lambda_{100} [H_{3}^{\dagger} H_{3}]_{1_{1}} [H_{3}^{'' \dagger} H_{3}^{''}]_{1_{1}} +
\lambda_{101} ([H_{3}^{\dagger} H_{3}^{''}]_{1_{1}} [H_{3}^{\dagger} H_{3}^{''}]_{1_{1}} +h.c.)\\+
\lambda_{102} [H_{3}^{\dagger} H_{3}^{'' \dagger}]_{1_{1}} [H_{3} H_{3}^{''}]_{1_{1}} +
\lambda_{103} ([H_{3}^{\dagger} H_{3}^{\dagger}]_{1_{1}} [H_{3}^{''} H_{3}^{''}]_{1_{1}} +h.c.)\\+
\lambda_{104} [H_{3}^{\dagger} H_{3}]_{1_{2}} [H_{3}^{'' \dagger} H_{3}^{''}]_{1_{3}} +
\lambda_{105} ([H_{3}^{\dagger} H_{3}^{''}]_{1_{2}} [H_{3}^{\dagger} H_{3}^{''}]_{1_{3}} +h.c.)\\+
\lambda_{106} [H_{3}^{\dagger} H_{3}^{'' \dagger}]_{1_{2}} [H_{3} H_{3}^{''}]_{1_{3}}+
\lambda_{107} ([H_{3}^{\dagger} H_{3}^{\dagger}]_{1_{2}} [H_{3}^{''} H_{3}^{''}]_{1_{3}} +h.c.)\\+
\lambda_{106}^{'} [H_{3}^{\dagger} H_{3}^{'' \dagger}]_{1_{3}} [H_{3} H_{3}^{''}]_{1_{2}} +
\lambda_{107}^{'} ([H_{3}^{\dagger} H_{3}^{\dagger}]_{1_{3}} [H_{3}^{''} H_{3}^{''}]_{1_{2}} +h.c.)\\+
\lambda_{108} ([H_{3}^{\dagger} H_{3}]_{3_{1}} [H_{3}^{'' \dagger} H_{3}^{''}]_{3_{1}} +h.c.)+
\lambda_{109} ([H_{3}^{\dagger} H_{3}^{''}]_{3_{1}} [H_{3}^{\dagger} H_{3}^{''}]_{3_{1}} +h.c.)\\+
\lambda_{110} [H_{3}^{\dagger} H_{3}]_{3_{1}} [H_{3}^{'' \dagger} H_{3}^{''}]_{3_{2}} +
\lambda_{111} ([H_{3}^{\dagger} H_{3}^{''}]_{3_{1}} [H_{3}^{\dagger} H_{3}^{''}]_{3_{2}} +h.c.)\\+
\lambda_{112} [H_{3}^{\dagger} H_{3}^{'' \dagger}]_{3_{1}} [H_{3} H_{3}^{''}]_{3_{2}}+
\lambda_{113} ([H_{3}^{\dagger} H_{3}^{\dagger}]_{3_{1}} [H_{3}^{''} H_{3}^{''}]_{3_{2}} +h.c.)\\+
\lambda_{114} [H_{3}^{' \dagger} H_{3}^{'}]_{1_{1}} [H_{3}^{'' \dagger} H_{3}^{''}]_{1_{1}} +
\lambda_{115} ([H_{3}^{' \dagger} H_{3}^{''}]_{1_{1}} [H_{3}^{' \dagger} H_{3}^{''}]_{1_{1}} +h.c.)\\+
\lambda_{116} [H_{3}^{' \dagger} H_{3}^{'' \dagger}]_{1_{1}} [H_{3}^{'} H_{3}^{''}]_{1_{1}} +
\lambda_{117} ([H_{3}^{' \dagger} H_{3}^{' \dagger}]_{1_{1}} [H_{3}^{''} H_{3}^{''}]_{1_{1}} +h.c.)\\+
\lambda_{118} [H_{3}^{' \dagger} H_{3}^{'}]_{1_{2}} [H_{3}^{'' \dagger} H_{3}^{''}]_{1_{3}} +
\lambda_{119} ([H_{3}^{' \dagger} H_{3}^{''}]_{1_{2}} [H_{3}^{' \dagger} H_{3}^{''}]_{1_{3}} +h.c.)\\+
\lambda_{120} [H_{3}^{' \dagger} H_{3}^{'' \dagger}]_{1_{2}} [H_{3}^{'} H_{3}^{''}]_{1_{3}} +
\lambda_{121} ([H_{3}^{' \dagger} H_{3}^{' \dagger}]_{1_{2}} [H_{3}^{''} H_{3}^{''}]_{1_{3}} +h.c.)\\+
\lambda_{120}^{'} [H_{3}^{' \dagger} H_{3}^{'' \dagger}]_{1_{3}} [H_{3}^{'} H_{3}^{''}]_{1_{2}} +
\lambda_{121}^{'} ([H_{3}^{' \dagger} H_{3}^{' \dagger}]_{1_{3}} [H_{3}^{''} H_{3}^{''}]_{1_{2}} +h.c.)\\+
\lambda_{122} ([H_{3}^{' \dagger} H_{3}^{'}]_{3_{1}} [H_{3}^{'' \dagger} H_{3}^{''}]_{3_{1}} +h.c.)+
\lambda_{123} ([H_{3}^{' \dagger} H_{3}^{''}]_{3_{1}} [H_{3}^{' \dagger} H_{3}^{''}]_{3_{1}} +h.c.)\\+
\lambda_{124} [H_{3}^{' \dagger} H_{3}^{'}]_{3_{1}} [H_{3}^{'' \dagger} H_{3}^{''}]_{3_{2}} +
\lambda_{125} ([H_{3}^{' \dagger} H_{3}^{''}]_{3_{1}} [H_{3}^{' \dagger} H_{3}^{''}]_{3_{2}} +h.c.)\\+
\lambda_{126} [H_{3}^{' \dagger} H_{3}^{'' \dagger}]_{3_{1}} [H_{3}^{'} H_{3}^{''}]_{3_{2}}+
\lambda_{127} ([H_{3}^{' \dagger} H_{3}^{' \dagger}]_{3_{1}} [H_{3}^{''} H_{3}^{''}]_{3_{2}} +h.c.)
\end{gather*}
\normalsize

\newpage

\section*{Acknowledgments}
This work was supported by DOE, grant number DE-FG02-05ER41418, and DOE-GAANN, award number
P200A090135.


\begin{thebibliography}{100}


\bibitem{Zwicky}
 F.~Zwicky,
 Helv.\ Phys.\ Acta {\bf 6}, 110 (1933).

\bibitem{Perlmutter}
 S.~Perlmutter {\it et al.},
 Astrophys.\ J.\ {\bf 517}, 565 (1999).
 {\tt arXiv:astro-ph/9812133}.
 
\bibitem{Riess}
 A.G.~Riess {\it et al.},
 Astron.\ J.\ {\bf 116}, 1009 (1998).
 {\tt arXiv:astro-ph/9805201}.

\bibitem{Goldberg}
 H.~Goldberg,
 Phys.\ Rev.\ Lett.\ {\bf 50}, 1419 (1983).

\bibitem{K}
 J.E.~Kim,
 Phys.\ Rev.\ Lett.\ {\bf 43}, 103 (1979).

\bibitem{Z}
 A.R.~Zhitnitsky,
 Sov.\ J.\ Nucl.\ Phys.\ {\bf 31}, 260 (1980).

\bibitem{SVZ}
M.A.~Shifman, A.I.~Vainshtein, and V.I.~Zakharov,
Nucl.\ Phys.\ B {\bf 166}, 493 (1980).

\bibitem{DFS}
 M.~Dine, W.~Fischler, and M.~Srednicki,
 Phys.\ Lett.\ B {\bf 104}, 199 (1981).

\bibitem{FKTY}
P.H.~Frampton, M.~Kawasaki, F.~Takahashi, and T.T.~Yanagida,
JCAP {\bf 1004}, 023 (2010).
{\tt arXiv:1001.2308 [hep-ph]}.

\bibitem{Carr}
B.J.~Carr, K.~Kohri, Y.~Sendouda, and J.~Yokoyama,
Phys.\ Rev.\ D {\bf 81}, 104019 (2010).
{\tt arXiv:0912.5297 [astro-ph]}.

\bibitem{NFW}
 J.F.~Navarro, C.S.~Frenk, and S.D.M.~White,
 Astrophys.\ J.\ {\bf 462}, 563 (1996).
 {\tt arXiv:astro-ph/9508025}.

\bibitem{Conf}
P.H.~Frampton,
Mod.\ Phys.\ Lett.\ A {\bf 22}, 931 (2007).
{\tt arXiv:astro-ph/0607391}.

\bibitem{FKtprime}
  P.H.~Frampton and T.W.~Kephart,
  Int.\ J.\ Mod.\ Phys.\ A {\bf 10}, 4689 (1995).
  {\tt arXiv:hep-ph/9409330}.

\bibitem{FK2001}
 P.H.~Frampton and T.W~Kephart,
 Phys.\ Rev.\ D {\bf 64}, 086007 (2001).
 {\tt arXiv:hep-th/0011186}.

\bibitem{CM}
 M.-C.~Chen and K.T.~Mahanthappa,
 Phys.\ Lett.\ B {\bf 652}, 34 (2007).
 {\tt arXiv:0705.0714 [hep-ph]}.
 
\bibitem{Feruglio:2007uu} 
F.~Feruglio, C.~Hagedorn, Y.~Lin, and L.~Merlo, 
Nucl.\ Phys.\  B {\bf 775}, 120 (2007); 
[Erratum-ibid.\  {\bf 836}, 127 (2010)]. 
{\tt arXiv:hep-ph/0702194}.  

\bibitem{Frampton:2007et}
  P.H.~Frampton and T.W.~Kephart,
  JHEP {\bf 0709}, 110 (2007).
  {\tt arXiv:0706.1186 [hep-ph]}.

\bibitem{Frampton:2008bz}
  P.H.~Frampton, T.W.~Kephart, and S.~Matsuzaki,
  Phys.\ Rev.\  D {\bf 78}, 073004 (2008).
  {\tt arXiv:0807.4713 [hep-ph]}.

\bibitem{FKR}
P.H.~Frampton, T.W.~Kephart, and R.M.~Rohm,
Phys.\ Lett.\ B {\bf 679}, 478 (2009).
{\tt arXiv:0904.0420 [hep-ph]}.

\bibitem{EFM1}
D.A.~Eby, P.H.~Frampton, and S.~Matsuzaki,
Phys.\ Lett.\ B {\bf 671}, 386 (2009). 
{\tt arXiv:0810.4899 [hep-ph]}.

\bibitem{EFM2}
D.A.~Eby, P.H.~Frampton, and S.~Matsuzaki,
Phys.\ Rev.\ D {\bf 80}, 053007 (2009). 
{\tt arXiv:0907.3425 [hep-ph]}.

\bibitem{FHKM}
  P.H.~Frampton, C.M.~Ho, T.W.~Kephart, and S.~Matsuzaki,
  Phys.\ Rev.\  D {\bf 82}, 113007 (2010).
  {\tt arXiv:1009.0307 [hep-ph]}.

\bibitem{SuperK}
 Y.~Fukuda et al.,
 Phys.\ Rev.\ Lett.\ {\bf 81}, 1158 (1998);
 [Erratum-ibid.\ {\bf 81}, 4279 (1998)].
 {\tt arXiv:hep-ex/9805021}.

\bibitem{Aranda:2000tm} 
A.~Aranda, C.D.~Carone, and R.F.~Lebed, 
Phys.\ Rev.\  D {\bf 62}, 016009 (2000). 
{\tt arXiv:hep-ph/0002044}.  

\bibitem{He:2003rm}
  X.G.~He and A.~Zee,
  Phys.\ Lett.\  B {\bf 560}, 87 (2003).
  {\tt arXiv:hep-ph/0301092}.

\bibitem{Babu:2005se}
  K.S.~Babu and X.G.~He,
  (2005).
  {\tt arXiv:hep-ph/0507217}.

\bibitem{Lee:2006pr}
  T.D.~Lee,
  Chinese Phys.\ {\bf 15}, 1125 (2006).
  {\tt arXiv:hep-ph/0605017}.

\bibitem{Frampton:1993wu}
  P.H.~Frampton, P.I.~Krastev, and J.T.~Liu,
  Mod.\ Phys.\ Lett.\ A {\bf 9}, 761 (1994).
  {\tt arXiv:hep-ph/9308275}.

\bibitem{HPS}
P.F.~Harrison, D.H.~Perkins, and W.G.~Scott,
Phys.\ Lett.\  B {\bf 530}, 167 (2002).
{\tt arXiv:hep-ph/0202074}.

\bibitem{Ma}
  E.~Ma and G.~Rajasekaran,
  Phys.\ Rev.\  D {\bf 64}, 113012 (2001).
  {\tt arXiv:hep-ph/0106291}.

\bibitem{Altarelli}
 G.~Altarelli and F.~Feruglio,
 Nucl.\ Phys.\ B {\bf 741}, 215 (2006).
 {\tt arXiv:hep-ph/0512103}.

\bibitem{Valle1}
 M.~Hirsch, S.~Morisi, E.~Peinado, and J.W.F.~Valle,
 Phys.\ Rev.\ D {\bf 82}, 116003 (2010).
 {\tt arXiv:1007.0871 [hep-ph]}.

\bibitem{Valle2}
 M.S.~Boucenna, M.~Hirsch, S.~Morisi, E.~Peinado, M.~Taoso, and J.W.F.~Valle,
 JHEP {\bf 1105}, 037 (2011).
 {\tt arXiv:1101.2874 [hep-ph]}.

\bibitem{TW}
A.D.~Thomas and G.V.~Wood,
{\it Group Tables},
(Shiva Publishing Ltd., Orpington, 1980).

\bibitem{FY}
M.~Fukugita and T.~Yanagida,
Phys.\ Lett.\ B {\bf 174}, 45 (1986).

\bibitem{strumia}
M.~Cirelli, N.~Fornengo, and A.~Strumia,
Nucl.\ Phys.\ B {\bf 753}, 178 (2006).
{\tt arXiv:hep-ph/0512090}.

\bibitem{WMAP}
N.~Jarosik, {\it et al.},
Astrophys.\ J.\ Suppl.\ {\bf 192}, 14 (2011).
{\tt arXiv:1001.4744 [astro-ph]}.

\bibitem{pdg}
K.~Nakamura {\it et al.} (Particle Data Group),
J.\ Phys.\ G {\bf 37}, 075021 (2010).

\bibitem{theta13}
H.~De Kerret,
Talk at LowNu Nov. 2011, Seoul National University.
{\tt http://workshop.kias.re.kr/lownu11/}

\bibitem{SV}
J.~Schechter and J.W.F.~Valle,
Phys.\ Rev.\ D {\bf 22}, 2227 (1980).

\bibitem{Mink}
P.~Minkowski,
Phys.\ Lett.\ B {\bf 67}, 421 (1977).

\bibitem{Frampton:2008ci}
  P.H.~Frampton and S.~Matsuzaki,
  (2008).
  {\tt arXiv:0806.4592 [hep-ph]}.
  
\end{thebibliography}
\end{document}